\documentclass[graybox]{svmult}

\usepackage{mathptmx}       
\usepackage{helvet}         
\usepackage{courier}        
\usepackage{type1cm}        
\usepackage{makeidx}         
\usepackage{graphicx}        
\usepackage{multicol}        
\usepackage[bottom]{footmisc}

\usepackage{amsmath}
\usepackage{float}

\usepackage{comment}
\usepackage{outlines}
\usepackage[usenames, dvipsnames]{xcolor} 

\usepackage[colorlinks]{hyperref}
\usepackage[colorinlistoftodos]{todonotes}

\usepackage{soul}

\usepackage{makecell}
\usepackage{booktabs}
\newcommand{\tabitem}{~~\llap{\textbullet}~~}
\usepackage{tabularx}
\newcolumntype{Y}{>{\centering\arraybackslash}X}
\usepackage{array}
\setcellgapes{4pt}

\makeindex             

\begin{document}

\title*{Gene regulatory networks: a primer in biological processes and statistical modelling} 
\titlerunning{Gene regulation - processes and modelling}
\author{Olivia Angelin-Bonnet, Patrick J. Biggs and Matthieu Vignes} 
\authorrunning{O. Angelin-Bonnet, P.J. Biggs, M. Vignes}
\institute{Olivia Angelin-Bonnet, Patrick J. Biggs \& Matthieu Vignes \at Institute of Fundamental Sciences\\
Patrick J. Biggs \at School of Veterinary Science \\ 
Massey University, Private Bag 11 222, Palmerston North, 4442, New Zealand\\ \email{m.vignes@massey.ac.nz}}

\maketitle

\abstract*{Each chapter should be preceded by an abstract (10--15 lines long) that summarizes the content. The abstract will appear \textit{online} at \url{www.SpringerLink.com} and be available with unrestricted access. This allows unregistered users to read the abstract as a teaser for the complete chapter. As a general rule the abstracts will not appear in the printed version of your book unless it is the style of your particular book or that of the series to which your book belongs.
Please use the 'starred' version of the new Springer \texttt{abstract} command for typesetting the text of the online abstracts (cf. source file of this chapter template \texttt{abstract}) and include them with the source files of your manuscript. Use the plain \texttt{abstract} command if the abstract is also to appear in the printed version of the book.}
\abstract{Modelling gene regulatory networks not only requires a thorough understanding of the biological system depicted but also the ability to accurately represent this system from a mathematical perspective. Throughout this chapter, we aim to familiarise the reader with the biological processes and molecular factors at play in the process of gene expression regulation. We first describe the different interactions controlling each step of the expression process, from transcription to mRNA and protein decay. In the second section, we provide statistical tools to accurately represent this biological complexity in the form of mathematical models. Amongst other considerations, we discuss the topological properties of biological networks, the application of deterministic and stochastic frameworks and the quantitative modelling of regulation. We particularly focus on the use of such models for the simulation of expression data that can serve as a benchmark for the testing of network inference algorithms.\\}



\section{Introduction}

\begin{figure}[htpb]
  \begin{center}
    \includegraphics[width=\textwidth]{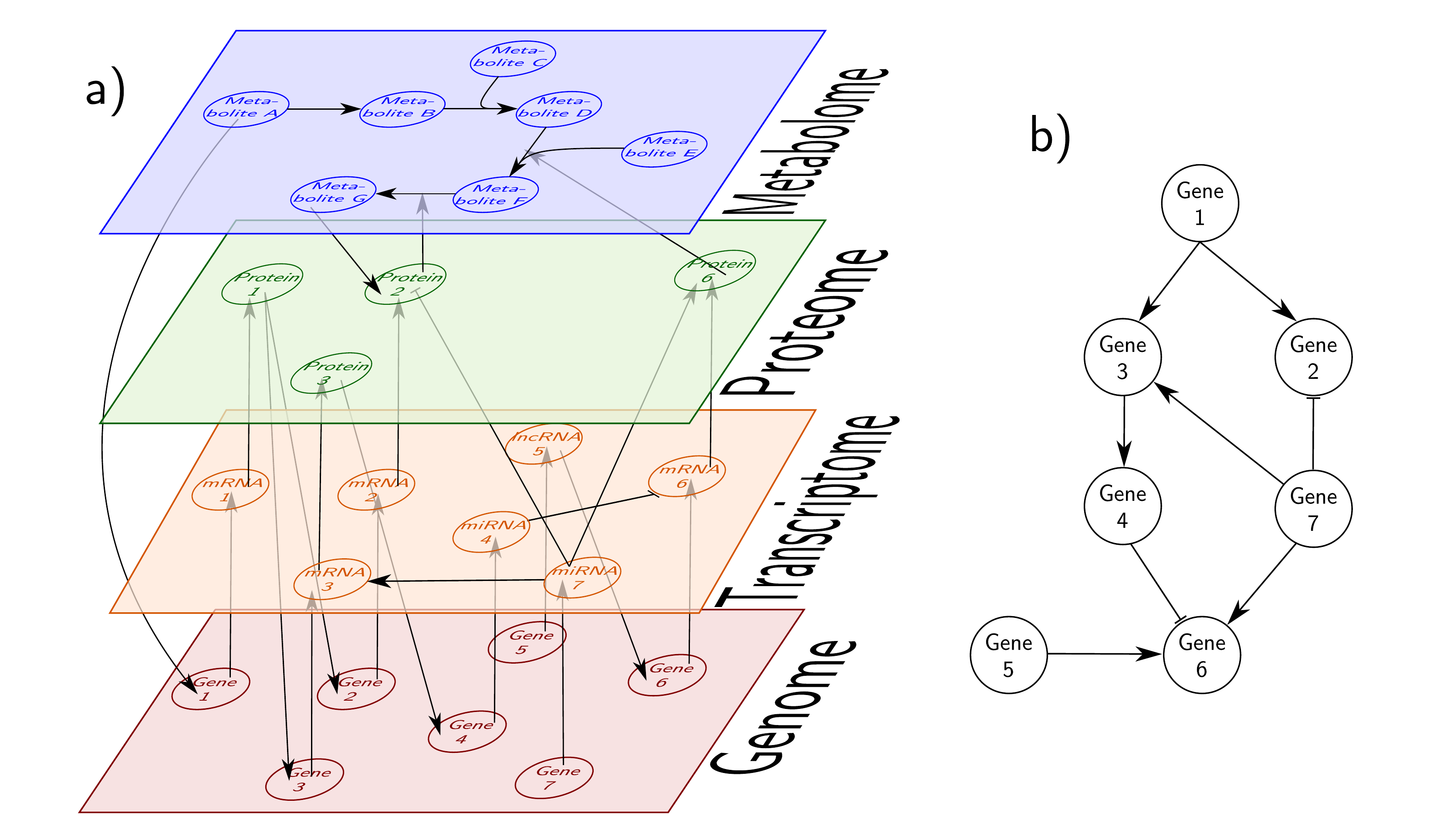}
   \caption{Biological versus statistical representation of a GRN. a) Biological regulatory systems are complex: the different intermediary products of genes --  transcripts and proteins -- as well as metabolites interact in a multi-layer network. Such networks are the best representation we can give of a biological complex system. b) Statistical perspective: genes can be considered as nodes in a directed graph, where the edges represent regulatory interactions. Each parent variable node directly influences its children variables, therefore representing the regulation mechanism of a gene product on the transcription of another gene.}  
  \end{center}
 \label{fig::GRN}
\end{figure}

The different regulatory processes occurring within cells are often depicted as a network of interacting entities. These entities can be mapped onto different layers that represent the different biological molecules involved in expression regulation, for example transcripts and proteins (Figure \ref{fig::GRN}a). High-throughput studies provide us with a measurement of the variable levels of a given layer. For example microarrays or RNA sequencing technologies measure mRNA abundance, and are commonly referred to as gene expression data. We refer the interested reader to~\cite{Conesa2016AAnalysis} for such modern data handling practices, to~\cite{Auer2010StatisticalData} for associated statistical designs and to~\cite{Backman2016SystemPipeR:Environment} for a data processing and primary analysis workflow.

From a biological perspective, entities from different layers are found to interact. Indeed, in addition to the well-known control of transcription by proteins termed transcription factors (TFs), other steps of the gene expression process are targeted by regulatory molecules beyond proteins, e.g. small molecules such as metabolites and noncoding RNAs. On top of this dynamic regulation, the information encoded in the DNA itself exerts to some extent control over the expression profile of genes. Here the term ``gene'' refers to a DNA sequence coding for a  protein or other untranslated RNA. However it is usually impossible to measure in the same experiment data about all these molecular layers. We are therefore most of the time bound to making the most of one given data type from which we seek to extract patterns giving insight into the regulatory interactions at play. Thus gene regulatory networks (GRNs) successfully gather the detected relationships between transcripts, even if these relationships are mediated by other molecules such as proteins. GRNs represent these interactions in a graph where nodes correspond to genes (and gene products) and edges represent the regulatory relationships among them (Figure \ref{fig::GRN}b).

The modelling of such regulatory systems is an important aspect of the reverse engineering problem. Accounting for existing biological interactions can be key to a more accurate analysis of experimental data, e.g. in the analysis of differential gene expression~\cite{Dona2017PowerfulData}. In addition, such models can be used to simulate expression data in order to assess the performances of a given network inference method, just like data can be simulated to assess gene expression differential analysis method performance~\cite{Rigaill2016SyntheticAnalysis}. Indeed, a detailed analysis of the strengths and weaknesses of a given method can guide the choice of a practitioner to choose among the possible different reverse engineering approaches and pave the way for needed method development. 
A possibility is to use as a benchmark a previously studied experimental dataset, but this approach is limited by our incomplete knowledge about true -- if this truth is ever an achievable objective -- underlying pathways. On the contrary, the use of simulated expression data from \textit{in silico} networks  renders possible the objective comparison of the results of network inference to the true underlying interaction graph. More precisely, synthetic data allows the assessment of the impact of sample size, noise or topological properties of the underlying network on the methods performance. To make valid conclusions, one expects synthetic data to have features as close as possible to real data. Modelling such complex systems seems like a insurmountable task. However, by carefully designing each constituting element of the model it is possible to link the statistical representation of a regulatory system to the underlying biological mechanisms in a meaningful way. This is the very topic of this work.

This chapter aims at bringing together the biological and statistical representation of GRNs. In Section~\ref{section::bio-proc-gene-prot}, we provide an overview of the different regulatory mechanisms that shape the gene expression profiles. We focus on the different regulatory molecules that target each step of the expression process. In Section~\ref{section::mod-gen-expr}, we introduce the reader to the basic concepts necessary to the construction of a GRN model, from the topological properties shaping biological networks to the mathematical frameworks used for the dynamic simulation of expression data and the representation of regulation from a quantitative point of view. Together, this chapter provides a first guide to GRN modelling anchored in the biological reality of gene expression regulation. \\

\section{Biological processes: from gene to protein}
\label{section::bio-proc-gene-prot}

\begin{figure}[htpb]
  \begin{center}
    \includegraphics[width=0.9\textwidth]{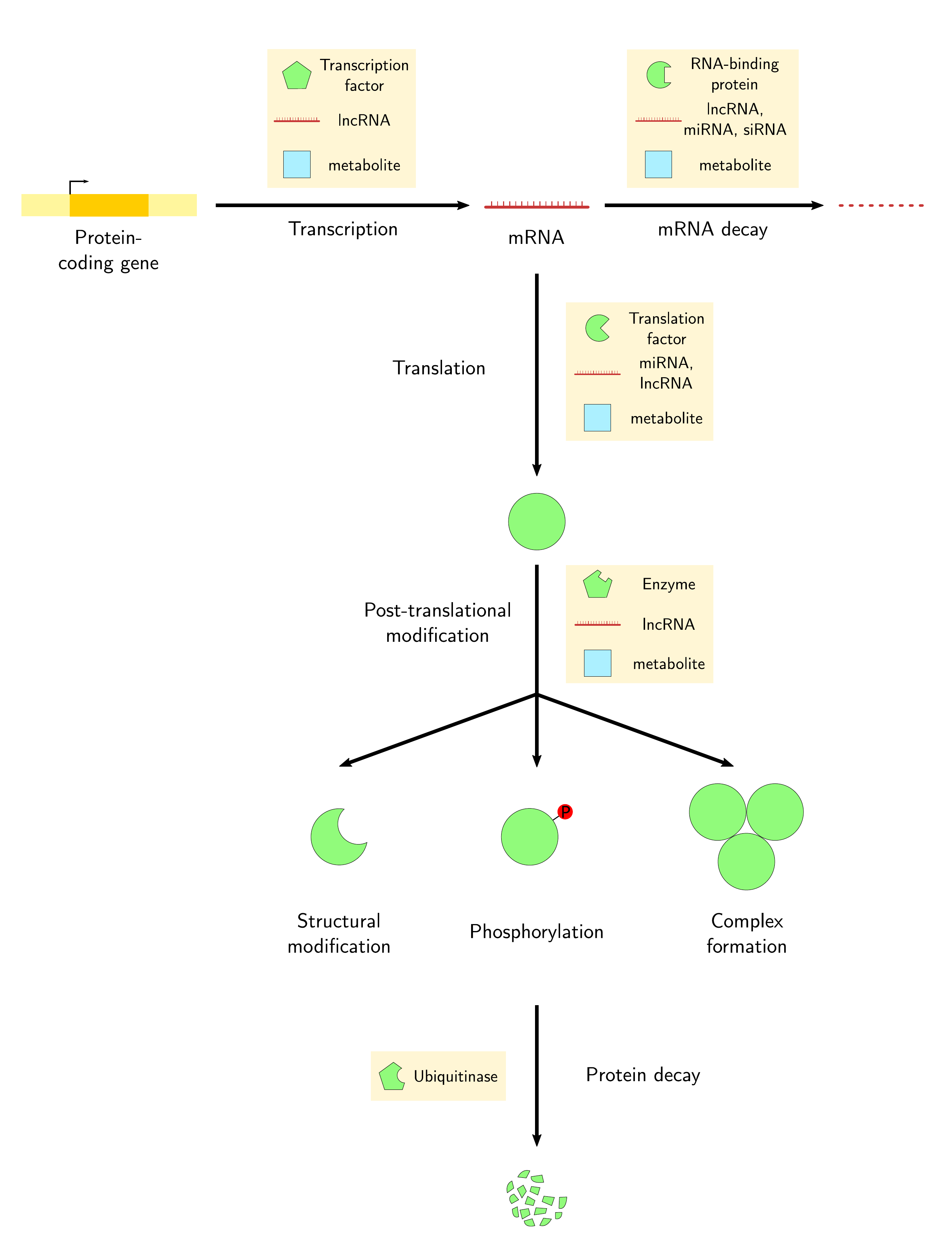}
   \caption{The different steps of the expression process of a protein-coding gene, and its possible regulatory molecules. The colors represent the different molecule types: yellow: DNA, red: RNA, green: protein, blue: metabolite. A gene is first transcribed into a mRNA, with the possible involvement of transcription factors, long noncoding RNAs (lncRNAs) or metabolites. The mRNA is then processed and translated into a protein; again this process can be affected by translation factors, microRNAs (miRNAs), lncRNAs, or small molecules. The degradation of transcripts is influenced by noncoding RNAs, RNA-binding proteins or metabolites. Once synthesised, a protein can undergo post-translational modifications, mediated by other proteins, lncRNAs or other small metabolites. Possible modifications include conformational change, modification of specific residues such as phosphorylation, or the formation of protein complexes. Proteins are tagged to degradation by specific enzymes, termed ubiquitinases.}  
   \label{fig::regulation_overview}
  \end{center}
\end{figure}

Proteins are the main actors in living organisms. They achieve a myriad of functions. Yet their structure, their production mechanisms and their regulation to allow the cell or organism to adequately adapt to the environment is dictated by the information contained in the genetic material of the organism.
The expression of a gene, a ``coding sequence'' into an active protein is a complex process involving numerous biological molecules transformed via varied reactions and interactions. The information encoded in the coding sequence of the DNA is transcribed into a messenger RNA (mRNA), which is processed and translated into a protein, according to the central dogma of biology. Once synthesized, a protein may require additional ``post-translational'' modifications to acquire a functional form. In this section, we aim at providing an overview of the different regulatory interactions targeting each of these steps. This knowledge certainly helps data analysts designing more \textit{ad hoc} models to extract knowledge from modern high-throughput measurements. While it is out of the scope of this chapter to provide a detailed and comprehensive description of the specific biological mechanisms, we provide references to more biology-centred reviews of the subject. An overview of the different molecular actors of this regulation can be found in Figure \ref{fig::regulation_overview}.\\

\subsection{Regulation of transcription}

The regulation of transcription is believed to be a key determinant of gene expression profile~\cite{Pai2015TheRegulation,Zlatanova2016MolecularProteomes}. It mainly leverages the action of TFs which act as activators or repressors for the transcription of target genes. 
Regulators act by binding to proximal or distant sites on the promoter of the target genes. They impact transcription by facilitating or restraining the recruitment of the transcriptional machinery to the target gene via protein-protein interactions with its constituents. While TF binding only involves proximal promoters in bacteria, additional remote regulatory elements such as enhancers, insulators or locus control regions play an important role in the regulation of eukaryotic genes~ \cite{Maston2006TranscriptionalGenome, Zlatanova2016MolecularProteomes}. A given TF can affect the expression of one or more target genes, and its impact on gene expression (i.e. activation, repression or modulation) can change in response to a specific environmental or molecular stimulus. Typically, a TF will only regulate a few targets, but some global TFs can control transcription of large sets of genes~\cite{Balaji2006ComprehensiveYeast}. Interestingly, while TFs play a crucial role in the control of gene expression, they are often found in low concentration, possibly only a few molecules per cell~\cite{Zlatanova2016MolecularProteomes}. 

Conversely, the transcription of a specific gene can be controlled by several TFs. This important feature, termed ``combinatorial regulation'', provides the cell with an increased complexity in transcriptional regulation. Each gene can potentially process several inputs which dictate its resulting expression profile~\cite{Balaji2006ComprehensiveYeast}. 
The different regulator molecules can act independently, if each of them affects a different aspect of the transcriptional machinery. Alternatively, TFs are often found to form complexes, either homo-dimers or hetero-dimers, thereby exerting cooperative regulation on the target~\cite{Balaji2006ComprehensiveYeast,Ravasi2010AnMan}. Importantly, such cooperation implies that the regulation only occurs when all the components of the regulatory complex are present. 
Yet another mechanism of combinatorial regulation is a synergistic interaction, where the global effect of the different TFs is greater than the sum of their individual effects~\cite{Maston2006TranscriptionalGenome,Schilstra2008Bio-Logic:Logic}. 
Finally, different TFs can compete for the same binding site on the target promoter. The respective affinity of the different molecules for the binding sequence determines which of them preferentially occupies the promoter. These affinities can be altered by environmental cues or changes in the promoter context (occupancy of neighbouring sites, etc). 

In addition to protein regulators, transcription can be controlled by noncoding RNAs, whose role will be discussed later in this chapter. Furthermore, recent evidence tend to suggest that small RNAs and in particular microRNAs also play a role in transcription silencing, in addition to their impact on post-transcriptional functions detailed in the following sections \cite{Castel2013RNABeyond, Catalanotto2016MicroRNAFunctions}. Lastly, other features can affect the transcription of genes. Specifically, the methylation state of DNA, and in particular of gene promoters, has been linked to gene silencing~\cite{Gonzalez-Zulueta1995MethylationSilencing, Herman2003GeneHypermethylation.}. A widespread example is the inactivation of tumour-suppressor genes by hypermethylation as a hallmark of cancer~\cite{Jones2007TheCancer}. DNA methylation is controlled by an array of specialized enzymes. 
In eukaryotic cells, the chromatin structure, that is the packaging of the DNA, affects all steps of the transcription process. This structure is dynamic and is regulated by ATP-dependent chromatin-remodelling complexes which control DNA-histones interactions, and by histone-modification factors~\cite{Li2007TheTranscription}. Histone post-translational modifications (such as methylation, acetylation, phosphorylation and more) greatly affect the chromatin structure, notably through the recruitment of chromatin-remodelling complexes, or by directly influencing their interactions with DNA. These histone marks have been found to correlate with transcription efficiency and in some case control the access of TFs to promoters \cite{Li2007TheTranscription}.

\subsection{Regulation of translation}

Following gene transcription and transcript processing, mRNAs are translated into proteins by ribosomes and associated molecules. This process is also targeted for regulation, both global and specific. The initiation of translation, that is binding of the translational apparatus to mRNAs and recognition of the translation starting site, is thought to be the step where most of the regulation occurs. As the specific mechanisms of translation initiation differ between bacteria and eukaryotes \cite{Kozak2005RegulationEukaryotes,Zlatanova2016MolecularProteomes}, regulation processes are specific to each, but similarities can be observed. Regulation of translation offers a faster modulation of the concentration of proteins compared to transcription regulation, as the former silences already existing mRNAs, while with the latter these mRNAs are still transcribed until their decay \cite{Sonenberg2009RegulationTargets}. 

As an example, during the response to a particular stress such as nutrient deprivation or temperature shock, cells often undergo a global decrease of their translational activity~\cite{Gebauer2004MolecularControl, Halbeisen2008Post-transcriptionalPrinciples, Sonenberg2009RegulationTargets}. This global programming switch occurs through the control of the availability or the activity of the translational apparatus, notably through the phosphorylation state of eukaryotic initiation factors in eukaryotes. Such massive translation reduction allows a decrease of the energy demand and a reallocation of cellular resources to stress response. Specific mRNAs encoding stress-response proteins can escape this regulation via distinct mechanisms. 

Alternatively to global programming, translation of mRNAs can be specifically regulated for a small set of genes via the involvement of RNA-binding proteins (RBPs) or microRNAs (miRNAs)~\cite{Gebauer2004MolecularControl, Merchante2017TranslationFuture}. These regulatory molecules recognise and bind to specific sequences in the target transcript, mainly situated in the untranslated regions of the mRNA. RBPs mainly act through interactions with the translational apparatus, leading to the inhibition of translation~\cite{Gebauer2004MolecularControl}. miRNAs act through the RNA-induced silencing complex (RISC) complex~\cite{Hutvagner2002AComplex, Valencia-Sanchez2006ControlSiRNAs}. The level of complementarity between a miRNA and its binding sequence on the target transcript specifies the triggered mechanism of regulation~\cite{Zlatanova2016MolecularProteomes}: the extensive base-pairing between the miRNA and its target triggers the degradation of the latter, whilst partial base-pairing induces translation inhibition~\cite{Jackson2007HowExpression}. Interestingly, it has been shown that in the case of miRNA-mediated translational repression, the promoter of the target gene determines the precise mechanism of action of the miRNA~\cite{Kong2008TheGene.}. However the role of small RNAs are still not perfectly clear and additional processes could be discovered by further experimental studies~\cite{Wu2008LetSiRNAs}.

It is interesting to note that the direct impact of small RNAs on the translation of a gene can also indirectly affect other processes such as transcription of non target genes. For example, \cite{Tu2009CombinatorialMechanisms} used miRNAs intervention experiments to detect their direct impact on TFs levels, but also reported the indirect effect of these miRNAs on the expression of theses TFs' targets. The conservation of miRNA-mRNA sequence match, particularly in the $ 3^{\prime} $ untranslated regions of genes enable the identification of the miRNA-target potential pairings~\cite{Friedman2009MostMicroRNAs}. Conversely, evidence suggest that miRNA synthesis can also be controlled by other RNAs~\cite{Guil2015RNA-RNAPlayers}. We can here again raise the concept of regulation network to start organising this knowledge. Prior interactions can be predicted to create such networks~\cite{Tu2009CombinatorialMechanisms,Wright2014CopraRNADomains}. Algorithms for RNA-RNA interaction predictions (e.g.~\cite{Salari2010FastInteraction}) are compared in~\cite{Lai2016AMethods.}. Then molecular techniques can confirm the putative relationships~\cite{Engreitz2014RNA-RNASites}.

The primary sequence of the transcripts also heavily influences their translation~\cite{Kozak2005RegulationEukaryotes}. In particular, the formation of secondary structures within the transcript (e.g. hairpin, stem-loop, etc. which can be facilitated by the properties of the primary sequence such as GC content for example), and specifically in regions involved in translation initiation can impair the translation process. Specific structural features, such as upstream open reading frames or internal ribosome entry sites can also impact translation. The detailed features of such mechanisms are beyond the scope of this chapter, and we refer the reader to~\cite{Kozak2005RegulationEukaryotes}. However, being aware of the existence of these mechanisms can allow the modeller to include them or at least discuss their effect on the outcome of an analysis. In this vein, \cite{Liang2007MicroRNANetwork.} postulate that protein-protein interactions are linked to the regulation of the corresponding genes by miRNAs.

Lastly, an interesting mechanism of translation control is the regulation via ``riboswitches''~\cite{Henkin2008RiboswitchMetabolism, Biggs2011RNARiboswitches, Serganov2012MetaboliteFunction}. A riboswitch is a regulatory sequence within mRNAs which responds to specific cues, namely temperature or the presence of particular metabolites. Thermo-sensors are a class riboswitches that respond to temperature by changing their conformation, therefore modifying the translation rate of the transcript. Alternatively, riboswitches can detect and link to specific metabolites. This provokes a modulated translational activity of the transcript via the modification of the mRNA conformation. \\

\subsection{Regulation of mRNA decay}

In addition to the elimination of defective mRNAs arising for example from transcription or splicing errors, fully functional mRNAs are subject to spontaneous or targeted degradation. Regulation of mRNA decay plays an important role in the resulting transcript level. Specific degradation of transcripts can be mediated by RBPs, or by small RNAs, namely miRNAs and small interfering RNAs (siRNAs)~\cite{Halbeisen2008Post-transcriptionalPrinciples}. Interestingly, mRNAs encoding functionally-related proteins were shown to exhibit correlated half-lives. This phenomenon suggests a common regulation of mRNAs involved in similar biological processes~\cite{Wang2002PrecisionDecay,Yang2003DecayAttributes}.

RBPs recognise and bind to specific sequences in mRNAs. It allows them to trigger the recruitment of decay factors, ultimately leading to target degradation. Alternatively, some RBPs have been found to stabilize their targets, protecting them from degradation~\cite{Kuwano2008MKP-1NF90}. Just as factors regulating other aspects of gene expression, RBPs can interact to exert combinatorial control over mRNA decay rates. 

Alternatively, miRNAs and siRNAs can promote the decay of target mRNAs, via interactions with RISC and possibly with other RBPs. Such a phenomenon is coined RNA interference or RNAi~\cite{Mattick2006Non-codingRNA., Valencia-Sanchez2006ControlSiRNAs}. As mentioned previously, such degradation is promoted by the perfect pairing of the small RNAs with the target transcript. Several mechanisms can be involved to trigger target degradation. Possibly, interactions with small RNAs and RISCs promote the endonucleolytic cleavage of the transcript. Another explanatory mechanism is that the target can be directed to P-bodies, which are small cytoplasmic granules containing RNA degradation machinery~\cite{Valencia-Sanchez2006ControlSiRNAs}. Targeted mRNAs are locked in these P-bodies and consequently degraded before they can be further processed, e.g. translated.

\subsection{Regulation of protein activity}

After translation, proteins are sometimes subjected to additional modifications to acquire their fully functional state. These changes can be irreversible, i.e. proteolytic cleavage of the peptidic precursor to obtain a functional protein~\cite{Cooper2000RegulationFunction, Zlatanova2016MolecularProteomes}. Alternatively, the cell can modulate the activity of its protein pool via a number of reversible post-translational modifications. 
A common mechanism is the modification by specialised enzymes of some amino acids on the protein, such as phosphorylation, oxidation or acetylation~\cite{Walsh2005ProteinDiversifications}. In particular, phosphorylation is a common mechanism for the activation of enzymes, TFs or other proteins. It is used in signalling pathways to relay extracellular messages to the nucleus, via a cascade of phosphorylation which activate kinase proteins~\cite{Hunter1995ProteinSignaling, Lizcano2002ThePathway}. The endpoints of such pathways are generally TFs, whose phosphorylation lead to their activation and relocalisation into the nucleus where they can modulate the expression of appropriate response genes. 

Taking again the example of signal transduction in the cell, the cascade of phosphorylation is initiated by the activation of membrane receptors, that detect a particular signal in the environment -generally a vitamin, a hormone, or another metabolite. This specific ligand binds to the receptor peptide, and triggers conformational changes to lead to the activation of the receptor. Such activation prompted by the binding of a small molecule is also frequently found in metabolic pathways, as a mean to regulate the production of a specific compound~\cite{Cooper2000RegulationFunction}. 
Metabolites can bind to the enzymes responsible for their synthesis in a feedback loop that auto-regulates enzyme activity according to the abundance of this specific product. Conformational changes resulting from ligand binding can mask or reveal the catalytic site of the enzyme, thereby controlling its ability to bind with its substrates. 

Lastly, peptidic chains sometimes need to assemble into multimers, to form a functionally active molecular complex~\cite{Cooper2000RegulationFunction}. Such protein complexes can be composed of several copies of the same protein, or of different proteins. In the latter case, the abundance of the complex, and hence its activity, is limited by the least abundant species. It is an interesting mechanism of regulation of the complex activity. Information about interactions among subunits can be found in protein-protein interaction databases (see for example~\cite{Szklarczyk2017TheAccessible}).

\subsection{Regulation of protein decay}

Cells possess several pathways for the degradation of proteins. A first mechanism is concerned with the degradation within lysosomes, which is a non-specific process, notably solicited in response to nutrient starvation as a rapid source of amino acids~\cite{Olson1989RegulationEukaryotes}. In addition, proteins can also be specifically tagged to degradation, via conjugation of a ubiquitin chain to the target peptide~\cite{Varshavsky2005RegulatedDegradation, Lecker2006ProteinStates,Zlatanova2016MolecularProteomes}. Tagged proteins are recognised by cellular machineries termed 26S proteasomes and subsequently degraded. This ubiquitin-proteasome pathway provides the cell with a way to rapidly control a regulatory process by degrading its effectors. It is notably involved in the regulation of transcription via degradation of specific TFs~\cite{Lecker2006ProteinStates}. 

The addition of ubiquitin on target proteins is mediated by the E1, E2 and E3 enzymes. The different isoforms of the E2 and especially E3 family confer a great specificity to this process, as each isoform can recognise different substrates. Additionally, some structural properties of proteins can impact their affinity as substrate for the ubiquitin-proteasome pathway. For example, a member of the E3 family recognises particular amino acids at the N-terminal position of proteins, in what is call the N-end rule pathway~\cite{Lecker2006ProteinStates}. The nature or accessibility of specific residues can also impact the ability of ubiquitination enzymes to recognise and tag target proteins. 

It is interesting to note that the ubiquitin-proteasome pathway is able to degrade only a subunit of a given protein, for example to produce a functionally active product or on the contrary inactivate the protein. This is the case for the NF-$\kappa$B TF, which is bound by its inhibitor, I$\kappa$B~\cite{Varshavsky2005RegulatedDegradation}. In response to a specific signal, the complex is ubiquitinated, and the proteasome cleaves the I$\kappa$B, thereby freeing the TF, which, in turn, is relocated into the nucleus to trigger the required cellular response. 

Quantitative measurements have highlighted the coupling between synthesis and decay rates of proteins. As for transcripts, these parameters seem to be correlated among proteins intervening in common complexes or functions. It appears that proteins involved in housekeeping functions are relatively stable, with a high production rate, leading to high concentrations in cells. On the contrary, regulatory proteins tend to be less synthesised and more rapidly degraded. This is consistent with the observation that they are often found in a low concentration in the cell~\cite{Belle2006QuantificationProteome, Vogel2012InsightsAnalyses}.

\subsection{The role of genetic variation}

In addition to diverse cellular molecules which perform a wide range of regulatory activities, the DNA sequence itself plays a role in regulating gene expression. Regulatory sequences present in the promoter region of genes or in the transcribed or translated sequences dictate the set of molecules and complexes that control the expression of these genes. These sequences target transcripts or corresponding proteins for particular regulatory mechanisms. Their affinity for regulators control the strength of this regulation. The impact of genetic variation on gene expression has been studied, notably via expression quantitative trait loci (eQTL) studies. eQTLs are genomic regions within which genetic variability is associated with variation in the abundance of a particular transcript~\cite{Gilad2008RevealingStudies}. More generally genetical genomics studies (also termed cellular genomics)~\cite{Jansen2001GeneticalSegregation,Gaffney2013GlobalVariation.} analyse how polymorphisms lead to variation in molecular traits, such as mRNA, protein or metabolite profiles. 

Using additional genomics data such as DNA methylation state or chromatin accessibility, researchers are now focusing on identifying the specific mechanisms which relate genetic variants to response molecular traits. 
At the transcript level, evidence tends to show that eQTLs lead to transcript abundance variability mainly via their impact on TF binding~\cite{Veyrieras2008High-resolutionRegulation, Gaffney2013GlobalVariation., Albert2015TheDisease}.  
Polymorphisms at these loci also affect other aspects of transcription, but it is yet to be determined if it is a direct consequence of genetic variation or merely an indirect effect of variation in TF binding efficiency~\cite{Pai2015TheRegulation}.  Some polymorphisms have also been shown to affect mRNA degradation, notably through modification of miRNA binding sites, or other post-translational mechanisms~\cite{Gaffney2013GlobalVariation., Pai2015TheRegulation}. In a groundbreaking effort, \cite{Bessiere2018ProbingCompositions} discovered instructions encoded in the sequence itself to regulate gene activity. The nucleotide composition can be directly read to accurately decipher biological mechanisms.

\subsection{An example: long noncoding RNAs}

After this review of the possible interactions regulating the different aspects of the gene expression process, we now turn our attention to a specific class of regulators whose role in the different biological processes mentioned earlier is just starting to be appreciated. Indeed, the functional importance of long noncoding RNAs (lncRNAs) was only hinted at when experimental studies of genome-wide transcription in cells revealed that a large fraction of the genome is transcribed, even if only a small amount actually encodes proteins~\cite{Rinn2012GenomeRNAs,Quinn2016UniqueFunction}. This discovery shook the traditional central dogma of biology stating that RNAs' primary role is to serve as messengers to produce functionally active proteins. On the contrary, as highlighted above, noncoding RNAs are now known to play important regulatory roles. While we now have a fair understanding of small noncoding RNAs (e.g. miRNAs, siRNAs, etc) and the associated biological processes, lncRNAs (exceeding 200 base-pairs, as an arbitrary defining threshold) remain for most of them \textit{terra incognita}. In particular, the extent of their functional role is yet to be determined, and there is still debate about whether the RNA molecule itself has a functional role or if only the physical changes triggered by its transcription (e.g. chromatin opening, helix unwinding, etc) impacts the transcription of neighbour genes while the produced transcript is useless~\cite{Wang2011MolecularRNAs, Zlatanova2016MolecularProteomes}. This is notably due to the fact that their primary sequence is less conserved than those of protein-coding genes~\cite{Mercer2009LongFunctions,Quinn2016UniqueFunction}. 
Nonetheless, experimental studies put us on the track of lncRNA involvement in a great variety of biological processes, from regulation of gene expression and chromatin state to genomic imprinting, in particular X chromosome silencing~\cite{Ponting2009EvolutionRNAs, Rinn2012GenomeRNAs, Quinn2016UniqueFunction}.
In addition, characteristic features of RNA make them well-suited for regulatory functions: their fast kinetics with no need for translation and their rapid degradation is particularly convenient for a fast and transient response to external stimuli. Moreover, their ability to bind DNA and RNA allows them to interact with both genes and transcripts~\cite{Wang2011MolecularRNAs,Geisler2013RNAContexts}. In this section, we briefly present the diverse roles played by lncRNAs in the regulation of gene expression. For a more thorough review about the biological roles of lncRNAs, we refer the reader to the review by Geisler \textit{et al.}~\cite{Geisler2013RNAContexts}. 

One of the primary focuses of early studies about lncRNAs was their involvement in chromatin modelling~\cite{Rinn2012GenomeRNAs}, ultimately resulting in the modulation of gene expression. lncRNAs can act as scaffold which bring together different chromatin-remodelling proteins and to assemble them into a functional complex~\cite{Wang2011MolecularRNAs,Rinn2012GenomeRNAs}. 
Alternatively, they can guide such proteins to a target location, triggering changes in chromatin structure~\cite{Mercer2009LongFunctions}. An interesting mechanism of action follows the transcription trail of the noncodingRNA which influences the chromatin state. It consequently impacts the transcription of neighbouring genes~\cite{Geisler2013RNAContexts}. 

lncRNAs can also enhance or repress the initiation of transcription via interactions with the basal transcriptional machinery. For example as a response to heat shock, the interaction of a specific lncRNA with RNA polymerase II triggers the inhibition of target genes~\cite{Mercer2009LongFunctions, Geisler2013RNAContexts}. Additionally, lncRNAs can target TFs and modulate their activity, by directly prompting conformational changes, by recruiting TFs onto the target promoter, or by withholding the TFs away from their targets~\cite{Zlatanova2016MolecularProteomes}. 

lncRNAs are also involved in other steps of the gene expression process. In particular, they can influence mRNA processing, in particular mRNA splicing and editing~\cite{Mercer2009LongFunctions, Geisler2013RNAContexts}. 
Some lncRNAs can impact translation efficiency of their target transcripts, through mechanisms which are still not totally clear~\cite{Geisler2013RNAContexts}. They also putatively control RNA stability, either by recruiting specialised degradation machinery to the transcript, or by competing with miRNAs for binding sites. In the latter case, lncRNAs play a protecting role for mRNAs in preventing or delaying their degradation. For example, they can lure miRNAs to competitively bind to the same targets~\cite{Wang2011MolecularRNAs, Geisler2013RNAContexts}. Finally, lncRNAs can assist in protein binding to modulate their activity. Target proteins can be TFs, chromatin remodellers, or other regulatory molecules. 

While a few well-studied lncRNAs provide evidence for a functional role of these transcripts, a lot remains unknown about them. In particular, it is important to keep in mind that the functional roles described above apply to a few number of characterized lncRNAs, and it is possible that a fraction of these transcribed noncoding genes are the result of transcriptional noise or experimental artefacts \cite{Ponting2009EvolutionRNAs}. The modeller has the choice to include such information for a few annotated lncRNAs only, or to include the different putative roles, e.g. in a Bayesian framework. \\

As demonstrated throughout this section, the expression of genes is subject to a tight regulation from which arises great biological complexity. We now embrace the point of view of the statistical modelling of such biological systems. In particular, we discuss the different aspects of the construction of a model that must be carefully thought out in order to faithfully describe the biological processes under study.

\section{Modelling gene expression}
\label{section::mod-gen-expr}

Statistical models of GRNs aim at reproducing biological systems from a mathematical perspective to permit, \textit{inter alia}, the simulation of their dynamical behaviour. A number of models for the simulation of expression data have been proposed in the last decades; an overview of the principal algorithms and their key features is presented in Table \ref{table::overview}. However, the design of such simulation tools is far from trivial. First and foremost, the model must be a faithful representation of the biological system, from the general topology of the underlying regulatory network to the quantitative regulation exerted on the genes and gene products. In addition, different mathematical frameworks can be used for the dynamic simulation of expression profiles, each of them carrying its own set of assumptions and limitations. 

\begin{figure}[ht]
  \begin{center}
    \includegraphics[width=0.9\textwidth]{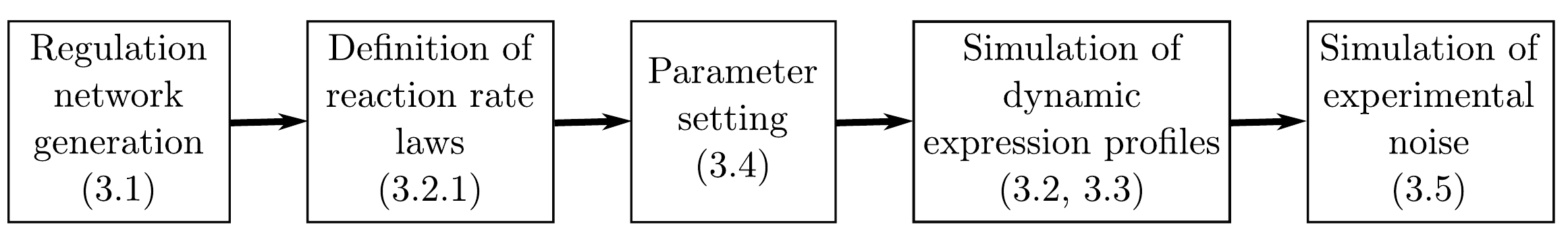}
   \caption{The different steps of an algorithm for expression data simulation. Each of these steps will be detailed in the referred sections.}  
  \end{center}
  \label{fig::pipeline_simulation}
\end{figure}

With all these considerations in mind, the next section provides a thorough reflection on the different features to examine when building a simulation network as well as a contrast of existing methods to simulate data from an \textit{in silico} network. The general pipeline for the construction of a simulation algorithm can be found in Figure \ref{fig::pipeline_simulation}. 

\subsection{Topological properties of regulatory networks} 

A crucial step in simulating expression data from regulatory networks is the selection of a network topology that defines the interactions among the molecules. In a graphical representation of a GRN, nodes typically represent genes and their products, while edges correspond to regulatory interactions between molecules. Edges carry the direction of the regulation, that is, which nodes are regulators and which nodes are target molecules. The choice of the topology of a GRN is by no means an easy task. A first and simple approach for network modelling is to represent regulatory networks as random networks \cite{Erdos1959OnGraphs,Kauffman1969MetabolicNets} (also termed Erd\"{o}s-R\'{e}nyi graphs) in which each pair of nodes has the same probability of being connected. This model was and is still used for expression data simulation. However topological analysis of pathways recovered from model organisms highlighted the existence of specific structural properties among biological networks, owing to the evolutionary constraints that shaped them. Algorithms for the generation of synthetic networks with similar properties where developed to construct more realistic models of biological systems. Interestingly, a number of these properties are shared with non biological systems such as the Internet or social networks \cite{Barabasi2004NetworkOrganization}:

\begin{itemize}
\item \textbf{Small-world property}: Networks are characterized as small-world if their average path length\footnote{The path length between a pair of nodes is defined as the length of the shortest path connecting the two nodes.} between any two nodes is small. It has been shown that most biological networks exhibit such a property (see for example \cite{Jeong2000TheNetworks,Wagner2001TheNetworks,Albert2007NetworkBiology.}). This implies that components of biological networks are easily reachable from any other node, which allows a rapid response to stimuli or perturbations \cite{Albert2007NetworkBiology.}. Synthetic random small-world networks are also referred to as Watts-Strogatz networks \cite{Watts1998CollectiveNetworks}.  \\

\item \textbf{Scale-free property}: When studying the in- and out-degree distribution of (directed) biological networks, i.e. the the number of incoming and outgoing edges respectively, it has been noted that this distribution can often be modelled with a power-law distribution~\cite{Barabasi1999EmergenceNetworks,Albert2000TopologyUniversality}. More specifically, the probability of a node to exhibit $k$ edges is $P(k) \propto k^{-\lambda}$. An implication is that the majority of nodes interact only with a few partners, while a small number of nodes, called hubs, are highly connected. Metabolic pathways were shown to have this property~\cite{Jeong2000TheNetworks,Wagner2001TheNetworks}, and so were GRNs~\cite{Featherstone2002WrestlingNetwork}. For both types of networks, the scale parameter $\lambda$ usually ranges between $ 2 $ and $ 3 $ \cite{Ravasz2002HierarchicalNetworks, Barabasi2004NetworkOrganization}. However recent findings suggest that for some organisms the in-degree distribution of transcriptional networks is not scale-free, as detailed below. An algorithm for generating random scale-free networks has been proposed by Albert-Barab\'{a}si \cite{Albert2000TopologyUniversality}. Bollob\'{a}s \cite{Bollobas2003DirectedGraphs} presented a directed version of scale-free networks, where both the in- and out-degree distributions are power laws, with possibly different $\lambda$ coefficients. \\

\item \textbf{Exponential distribution of the in-degree distribution} (for transcriptional networks): alternatively to the scale-free property, studies \cite{Guelzim2002TopologicalNetwork, Balaji2006ComprehensiveYeast} suggested that the in-degree distribution of GRNs for some organisms is better fitted by an exponential distribution, i.e. $P(k) \propto \frac{1}{\lambda}e^{-\frac{k}{\lambda}}$.  This implies that genes are regulated only by a few (generally up to three) TFs~\cite{Barabasi2004NetworkOrganization}, a more plausible configuration in biological networks.  \\

\item \textbf{Modularity}: real networks have a tendency to form groups of highly interconnected nodes, referred to as modules. This modular organization is characterized by a high average clustering coefficient \cite{Watts1998CollectiveNetworks,Wagner2001TheNetworks}. The clustering coefficient $ C $ of a node is a measure of the degree of connectivity among the direct neighbourhood of this gene. This property is important for biological systems, as it implies that biological networks are organised into relatively independent modules that each perform a distinct biological function. While inside a module the components are tightly linked, modules are only weakly connected with each other. This last property ensures to some degree robustness to the network, as disruption in one module is less likely to severely impair the rest of the network~\cite{Barabasi2004NetworkOrganization}. Methods to identify modules within pathways~\cite{Sanguinetti2008MMG:Pathways} could clearly inform gene network inference and this information should not be ignored when exploitable. \\

\item \textbf{Hierarchical organization}: contrary to random or scale-free networks for which the average clustering coefficient decreases with the number of nodes in the network, biological networks are characterized by a system-independent average clustering coefficient~\cite{Ravasz2002HierarchicalNetworks}. Moreover, the clustering coefficient of a node is a function of its degree~\cite{Ravasz2002HierarchicalNetworks}, since: $C(k)  \propto k^{-1}$.  This last property is a characteristic of a hierarchical organization of the network. This important mathematical concept allows us to reconcile the scale-free property and modular nature of biological systems. Indeed, it stresses that nodes of low connectivity tend to be found in clusters, while hub nodes constitute the junction between modules. It is to note that hub nodes will less likely be connected to each other.\\
 
\item \textbf{Over-representation of network motifs}: another important feature of biological networks is the abundance of small regulatory motifs~\cite{Milo2002NetworkNetworks,Shen-Orr2002NetworkColi}, that are recurring and non-random building blocks of the global topology~\cite{Zhu2005StructuralOrganisms}. They confer specific advantages to the system by encoding well-defined local dynamic behaviours in response to perturbations, for example buffering intrinsic stochasticity or on the contrary amplifying an external signal to trigger a cellular response \cite{Alon2007NetworkApproaches}. One well-known example is the negative feedback loop, in which the product of a gene regulates its own transcription \cite{Rosenfeld2002NegativeNetworks}. This auto-regulation feature allows the control of the natural fluctuation in the concentration of the gene product, as its synthesis is directly coupled to its abundance~\cite{Alon2007NetworkApproaches}. Another famous example is the feed-forward loop, who can simultaneously process two different stimuli and whose output depends on the nature (activation or repression) of the regulatory interactions composing the motif \cite{Mangan2003StructureMotif.}. A detailed quantitative analysis of such motifs and the advantages they provide to the system can be found in the book \cite{alon2006introduction}.\\
\end{itemize}

As pointed out by Przulj \textit{et al.}~\cite{Przulj2004ModelingGeometric}, such studies are based on our current and incomplete knowledge of biological networks. Despite this limitation, algorithms for the generation of graphs mimicking these structural properties have been proposed, for a more accurate representation of biological networks. In addition to the three most commonly used Erd\"{o}s-R\'{e}nyi, Albert-Barab\'{a}si and Watts-Strogatz networks, Haynes \textit{et al.}~\cite{Haynes2009BenchmarkingGRENDEL} implemented a method for simulating topologies with scale-free out-degree distribution and any desired in-degree distribution. Di Camillo \textit{et al.}~\cite{DiCamillo2009AAlgorithms} proposed a hierarchical modular topology model that generates networks displaying scale-free degree distribution, high clustering coefficient independent of the network size, and low average path length. However, to offer flexibility in the simulation, most simulators offer as an option for the user to choose among the different network topologies cited above. It becomes therefore possible to assess the impact of the underlying topological properties on the performances of a given network inference algorithm. 

The main drawback of \textit{in silico} networks is that none of the aforementioned network simulation methods are able to simultaneously reproduce all characteristic features of real networks~\cite{denBulcke2006SynTReN:Algorithms}. Another approach for graph generation has hence been proposed. It relies on the use of real biological networks determined experimentally. They are used as seeds from which sub-networks are sampled. Van den Bulcke \textit{et al.}~\cite{denBulcke2006SynTReN:Algorithms} proposed two sampling approaches: the cluster addition method and the neighbour addition method. Building up on this idea, Marbach \textit{et al.}~\cite{Marbach2009GeneratingMethods} further refined the approach by forcing the preferential inclusion of modules in the sampled sub-networks. This module extraction method ensures a fair representation of network motifs in the generated topology, as observed in biological networks. Such an approach of sampling from real networks ensures a more faithful picture of biological pathways. Again, this is contrasted by~\cite{denBulcke2006SynTReN:Algorithms}, as the real network sampling strategy relies on our current knowledge of regulatory networks, which is still incomplete, and probably biased towards well studied pathways. Lastly, Haynes \textit{et al.}~\cite{Haynes2009BenchmarkingGRENDEL} pointed out that networks generated from the same source network may not be ``statistically independent" as they may overlap and thus provide redundant information. It is particularly true when sampling a large number of subnetworks from a single source, as most of them will share common nodes and interactions.

\subsection{Mathematical frameworks and regulation functions}

Once the network topology is set, the next step for data simulation is to decide on the mathematical framework to be used to compute the profiles of gene expression, which will impact the choice of regulation rules for the system. It is important to carefully consider the different options, as each formalism carries a number of underlying assumptions about the represented system. Moreover, different levels of precision about the system can be integrated. Choices depend on many factors to achieve a balance between the level of required details to make the simulations more realistic, and the computational efficiency desired. While it is not our goal to offer an extensive comparison of all the possible formalisms, we emphasize here the difference between the two mainstream formalisms in existing simulators of expression data: the continuous-and-deterministic and discrete-and-stochastic frameworks. We present the basic concepts of these approaches, and highlight the different hypotheses about the biological system underlying each model. For a more detailed and mathematically-centred review of these and other formalisms we refer the reader to \cite{deJong2002ModelingReview.} and \cite{Higham2008ModelingReactions}.

\subsubsection{The continuous and deterministic approach}

The continuous and deterministic approach is particularly suited to simulate data that resemble those resulting from a transcriptomics (or other 'omics) experiment. The output is a series of continuous variables, as opposed to cruder logical models that predict the activation state of each gene as a binary outcome. 
In such deterministic models, biological molecules are represented as time-dependent continuous variables. Typically $x_i(t)$ represents the concentration of entity (or species) $i$ at time $t$. Variation in the concentration of a species over time is assumed to occur in a continuous and deterministic way. Such changes are modelled as differential equations, in the form of: \\
\begin{align}
\frac{dx_i}{dt} = f_i(\mathbf{X}),
\label{eq:RRE}
\end{align}
\noindent where $\mathbf{X}$ refers to the state of the system (that is the concentration of all the species present in the system), and $f_i$ represents the change in the concentration of species $i$ as a function, often non-linear, of the global state of the system. More specifically, in the case of a chemical species and associated reactions, $f_i(\mathbf{X})$ can be written as:
\begin{align}
f_i(\mathbf{X}) = v_i(\mathbf{X}) - d_i(\mathbf{X}),
\label{eq::chemical_reactions}
\end{align}
\noindent where the vector $v_i(\mathbf{X})$ represents the synthesis rate for species $i$, while $d_i(\mathbf{X})$ models its decay rate (due to degradation, dilution, use as a reactant, etc.). Both rates are themselves expressed as a function of the system state. The ensemble of reactions occurring in the network provides a set of coupled differential equations that describe the evolution of the state of the system through time. Except for simple networks, with only a few molecules and/or linear interactions, an analytical solution is often intractable. It is however possible to integrate the model in order to compute a numerical solution. A plethora of differential equation system solvers are available in different programming languages (for example the \texttt{deSolve} package for R, the \texttt{dsolve} function of the Symbolic Math toolbox for Matlab or Berkeley Madonna). 

Regulatory molecules for species $ i $ appear on the right-hand side of Equations~\ref{eq:RRE} and~\ref{eq::chemical_reactions}. They impact the abundance of their target species $ i $ via regulation of the different expression steps, as we discussed them in Section~\ref{section::bio-proc-gene-prot}.  The regulation of a particular reaction is hence modelled via a reaction rate law, which dictates how the rate of the regulated reaction (e.g. $v_i$ or $d_i$) evolves with the concentration of the different regulatory molecules. 
The vast majority of proposed simulation algorithms focus on the representation of transcription regulation, but similar consideration can be applied to any type of regulation (translation, degradation...). Two important features must be considered in order to fully characterize a reaction rate law: (i) the quantitative relation between a regulator abundance and the resulting reaction rate, and (ii) the combination scheme of the individual effects of different regulators on a common target. In the following we will discuss these two aspects, using the example of the regulation of gene transcription.

\begin{figure}[ht]
  \begin{center}
    \includegraphics[width=0.9\textwidth]{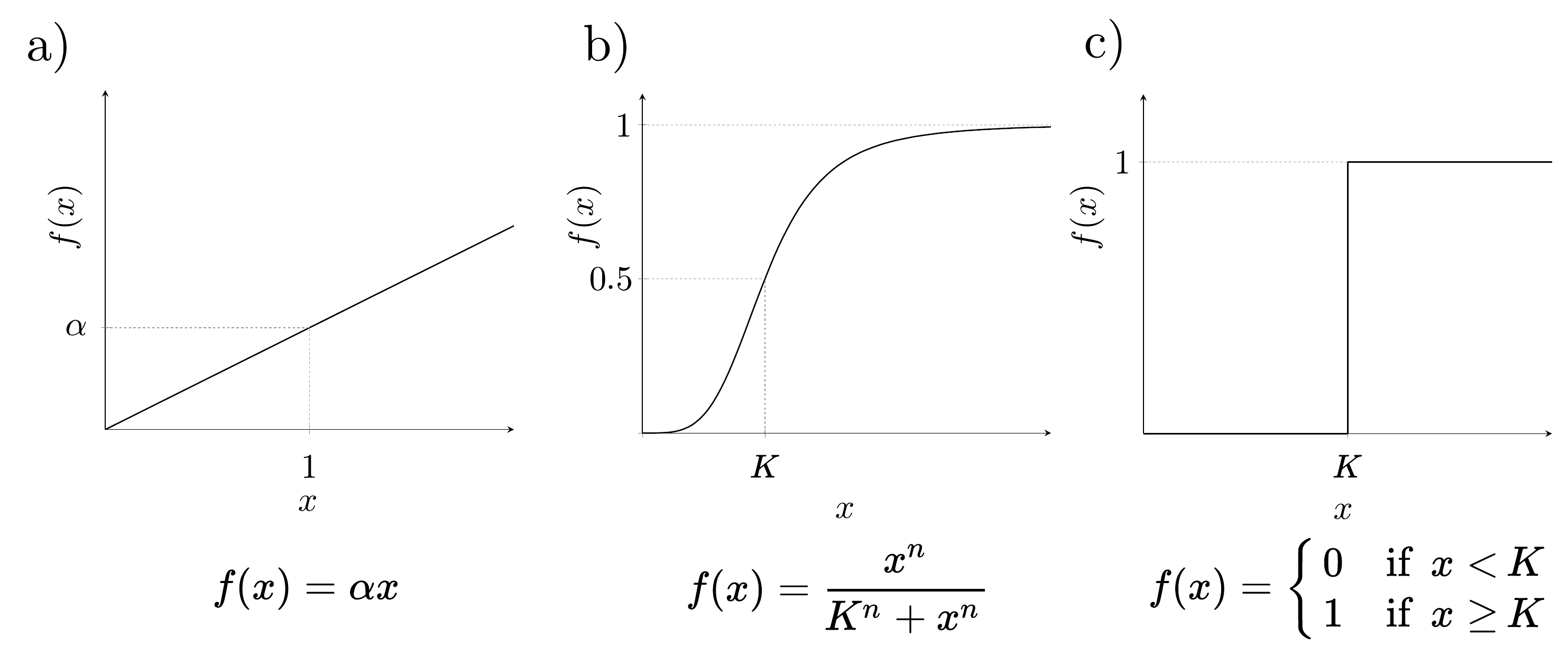}
   \caption{Possible transcription rate law functions. a) The rate law accounts for a linear dependence between the concentration of the regulator ($x$) and the transcription rate of the target gene ($f(x)$). b) The Hill function accounts for the saturation of the regulation: when $x$ is high, the variation of the transcription rate tends to $0$. The parameter $K$ corresponds to the concentration at which the regulatory molecules induce a transcription rate equal to half its maximum value. c) With a step function, the target gene is only transcribed when the concentration of the regulator exceeds a certain threshold, here $K$.}  
    \label{fig::reaction_rate_law}
    \end{center}
\end{figure}

We first consider gene $i$ whose transcription is regulated by a single molecule $j$. Several choices are possible with regard to the resulting effect of the regulator abundance on the transcription rate. A simple approach is to consider that the regulation effect increases linearly with the regulator concentration~\cite{Dhaeseleer1999LinearInjury.,Yeung2002ReverseRegression.} (Figure \ref{fig::reaction_rate_law} a), as follows: 
\begin{align}
f_i(x_j) = \beta \cdot x_j
\end{align}
where $f_i$ represents the transcription rate law of gene $i$, which depends on the concentration of the regulatory molecule $x_j$. In addition to the fact that this representation ignores any saturation effect arising from the limited amount of cellular resources and the maximum possible number of simultaneous transcription events, such a linear relationship can produce concentration values out of the plausible range of abundance encountered \textit{in vivo}, possibly leading to infinitely large populations, which is biologically irrelevant. It is hence important to construct biologically credible reaction rate laws that result in realistic regulation strength and concentration values.

Alternatively, a Hill function (Figure \ref{fig::reaction_rate_law} b) can be used to model the impact of an activator on the gene transcription:
\begin{align}
f_i(x_j) = \frac{x_j^{n_{ij}}}{x_j^{n_{ij}} + K_{ij}^{n_{ij}}},
\end{align}

or, for a repressor:

\begin{align}
f_i(x_j) = \frac{K_{ij}^{n_{ij}}}{x_j^{n_{ij}} + K_{ij}^{n_{ij}}},
\end{align}

\noindent where $K_{ij}$ represents the concentration of regulator $j$ required to obtain a half-maximum effect on the transcription rate of gene $i$, and $n_{ij}$ controls the steepness of the regulation. It must be noted that $K_{ij}$ must be non negative as it accounts for a concentration, and $ n_{ij} \geq 1 $. Indeed, when $n_{ij}=0$ the resulting reaction rate law is constant. This sigmoid function accounts for the saturation of the regulatory effect: after the regulator concentration has reached a certain level, any further increase in this concentration will only result in a minimal change in the transcription rate. Additionally, tuning the parameter $n_{ij}$ enables to represent a variety of regulation behaviours, from a quasi-linear ($n_{ij}$ small) to a step-like ($n_{ij}$ high) function. Furthermore the use of a Hill function law can be justified by a thermodynamic model of the binding of TFs on the target promoter~\cite{Ackers1982QuantitativeRepressor., Bintu2005TranscriptionalModels}. Considering that the mean transcription rate of a gene is proportional to the saturation of its promoter by TFs, the effect of the regulator can be further refined as:
\begin{align}
f_i(x_j) = \alpha_0 \left[ 1 + \left( FC_{ij} -1 \right) \frac{x_j^{n_{ij}}}{x_j^{n_{ij}} + K_{ij}^{n_{ij}}} \right] ,
\label{eq:TCratelawFC}
\end{align}
\noindent where $\alpha_0$ represents the basal transcription rate of the target gene in absence of the regulator, and $FC_{ij}$ the maximum fold-change\footnote{The fold-change of a gene is defined as the ratio of its transcription rate in the presence of a high concentration of regulatory molecules over its transcription rate in absence of regulator. From equation \ref{eq:TCratelawFC} it is easy to see that the transcription rate of gene $i$ tends towards $\alpha_0 \cdot FC_{ij}$ when $ x_j $ becomes large, hence the fold-change tends to $\frac{\alpha_0 \cdot FC_{ij}}{\alpha_0} = FC_{ij}$.} of gene expression induced by the regulator. Details of this computation can be found in the Supporting Information of \cite{Marbach2010RevealingInference}. It is straightforward to deduce the transcription rate law for a gene whose expression is controlled by an inhibitor, as the resulting maximum fold-change is $FC_i = 0$:
\begin{align}
f_i(x_j) = \alpha_0 \left[ 1 - \frac{x_j^{n_{ij}}}{x_j^{n_{ij}} + K_{ij}^{n_{ij}}} \right]
\end{align}
Such representation is massively used in simulator algorithms, although with some variations ~\cite{Mendes2003ArtificialAlgorithms,denBulcke2006SynTReN:Algorithms,Roy2008ATranscription,Haynes2009BenchmarkingGRENDEL,Hache2009ReverseStudy.,Schaffter2011GeneNetWeaver:Methods,Pinna2011SimulatingSysGenSIM}. For example, the transcription rate law represented in Equation~\ref{eq:TCratelawFC} can be adapted to account for a gene that is not expressed in the absence of its activator.

Taking the approximation that $n_{ij} \rightarrow \infty$, it is possible to simplify this Hill function, and model the transcription rate law as an on-off switch, where the maximum effect of the regulator on the transcription rate occurs as soon as the regulator molecule level exceeds a certain threshold. Below this threshold, no regulation is observed. Such representation is described by a step function (Figure \ref{fig::reaction_rate_law} c):
\begin{align*}
f_i(x_j) = \begin{cases}
					\alpha_0 \quad x_j < K_{ij} , \\
                    \alpha_1 \quad x_j \geq K_{ij}.
				\end{cases}
\end{align*}
$\alpha_0$ can be set to $0$, to model the case of a gene that is not expressed in absence of its regulator. This simplification provides the basis for piecewise differential equations \cite{deJong2002ModelingReview.}. It allows us to model a non-linear interaction even if the kinetics of the regulation are not known in detail. 

Once the quantitative effect of a regulator has been chosen, one must consider the overall effect of several regulators targeting a common gene. Indeed, different combinatorial regulations can be modelled. A simple example is to assume that different regulators impact the expression of the target gene independently of each other. This approach has been used by Mendes \textit{et al.}~\cite{Mendes2003ArtificialAlgorithms}. For an independent combinatorial effect model, the resulting regulation effect of all regulators is equal to the product\footnote{The use of the product, rather than the sum, ensures that if the concentration of a repressor is high enough to silence the gene (resulting in an individual effect close to $0$) the overall transcription rate will also tend to $0$ regardless of the quantity of activators present. It also implies that the overall fold-change obtained for large quantities of the different activators is the product of the fold-changes individually induced by each activator, which is justified thermodynamically in \cite{Bintu2005TranscriptionalApplications} and \cite{Bintu2005TranscriptionalModels}.} of the individual effects of each regulator on the transcription rate.
 
Alternatively, the different TFs can assemble into a complex that will bind to the target promoter to regulate transcription. In this case, the resulting regulation effect will be limited by the least abundant regulator species. An example can be found in \cite{Roy2008ATranscription}, in which different TFs can assemble into cliques which in turn can form TF complexes regulating the target gene. The resulting translation rate law is therefore equal to $ 0 $ as soon as the concentration of one of the TFs reaches $ 0 $, as it is then not possible to form a functional complex.
    
An interesting approach has been proposed by Di Camillo \textit{et al.}~\cite{DiCamillo2009AAlgorithms}. It uses fuzzy logic to represent the possible combinatorial interactions between different regulators. The advantage of such an approach is that it combines the Boolean logic functions (AND, OR, NOT, etc.) well suited to describe combinatorial behaviour with continuous regulation, as the output of fuzzy logic functions is a continuous value. Given a continuous input, that is the concentration of each regulator, the fuzzy logic applies a set of functions such as $ \min $, $ \max $, or $ \sum $ ($\mathrm{sum}$) to output the level of regulation commonly achieved by the different regulators. Di Camillo \textit{et al.} hence represents ``cooperation'' (for which the regulation is only achieved in the presence of all the required regulators) as a $ \min $ function applied to the set of regulator concentrations. Similarly, synergistic behaviour, direct inhibition or competition are modelled with fuzzy logic functions. 

Deterministic models are traditionally used for the simulation of expression data~\cite{Mendes2003ArtificialAlgorithms, denBulcke2006SynTReN:Algorithms, Roy2008ATranscription, DiCamillo2009AAlgorithms, Hache2009GeNGe:Networks, Haynes2009BenchmarkingGRENDEL, Marbach2010RevealingInference} (see Table \ref{table::overview}). However, despite its broad use, the deterministic formalism presents several limitations, which relates to the underlying hypotheses about the biological system depicted. In particular, the assumption of continuous change in species concentration is only valid for a macroscopic description of biological systems~\cite{deJong2002ModelingReview.}, i.e. when the number of molecules in the cell is large enough so that species concentrations can be considered to vary continuously when a discrete number of molecules is actually added/withdrawn from the system. When the abundance of a species reaches low values (defined as a thousand or less by \cite{Cao2009DiscreteSystems.}), this assumption does not hold anymore, and it is more correct to represent this abundance by a discrete molecule count. Moreover, the deterministic assumption can be questioned, particularly for small systems, given the fluctuation in the timing of biochemical reaction events \cite{deJong2002ModelingReview.}. Indeed two identical genes with the same transcription rate will not produce exactly the same number of transcripts during the same time period, due to the apparent stochasticity of biological events. While this fluctuation can be averaged out for highly abundant species, it is more difficult to ignore it for species with only a few molecules per cell. As numerous studies have underlined the importance of stochasticity in biochemical systems~\cite{ROSS1994TranscriptionInfrequently, McAdams1999ItsScale, Wilkinson2009StochasticSystems,Wilkinson2012StochasticBiology}, it can be preferable to explicitly model stochasticity in the simulation. \\

\subsubsection{The discrete and stochastic approach}

To overcome the limitation of continuous and deterministic models, in particular for modelling small systems, a discrete and stochastic representation of biological systems has been proposed. It must be noted that even if the continuous and deterministic approach and the discrete and stochastic framework are commonly referred to as respectively deterministic and stochastic models, there exists representations of biological systems that are either discrete and deterministic (e.g. Boolean networks) or continuous and stochastic (e.g. Chemical Langevin Equation, discussed later). One must hence keep in mind that continuous (resp. discrete) does not necessarily imply deterministic (resp. stochastic) as the first terms refer to the representation of species abundance while the latter corresponds to the variation of the system state. 

In the discrete and stochastic framework, the state of the system corresponds to discrete values accounting for the number of molecules of each species present in the system. While the vast majority of deterministic approaches are species-centred, i.e. one differential equation represents the evolution of one species abundance through time, stochastic models often rely directly on the biochemical reaction formalism. These reactions can be schematically represented in the form: 

\begin{align*}
\text{Substract 1} +\text{Substract 2} &\xrightarrow{\text{Reaction rate}} \text{Product 1} \\
\end{align*}

Or, in the context of gene expression:
\begin{align*}
\text{Promoter} + \text{TF} &\xrightarrow{c_1} \text{Active\_promoter} \\
\text{Active\_promoter} &\xrightarrow{c_2}  \text{Active\_promoter} + \text{mRNA} \\
\end{align*}
Each reaction is characterized by:
\begin{itemize}
\item A stoichiometry vector  $v_j$ which represents the change in abundance of the different species resulting from one firing (i.e. one occurence) of the reaction. Negative and positives indices correspond respectively to reactants and products of the reaction.
\item A propensity function $a_j(\mathbf{X})$, with $a_j(\mathbf{X}) \tau$ representing the probability that the reaction will occur in the next time step $[t, t+\tau]$ given the system state at this time $t$. The rate $a_j$ depends on the current state of the system. If $c_j$ is the constant probability that one molecule of each of the $r$ reactant species $S_i$, $1 \leq i \leq r$ collide and undergo the reaction in the next time unit, then $a_j = c_j.\prod \limits_{r} x_r$. Generally the number of reactants per reaction is limited to one or two, as a reaction involving more substrates can be decomposed into a set of elementary reactions~\cite{Gillespie2007StochasticKinetics}. Taking the example reactions above, the propensity function of the binding reaction of one TF molecule on the promoter will be: $c_2 \cdot x_{Promoter}x_{TF}$. It is therefore possible to link deterministic and stochastic rate constants, as shown in~\cite{Gillespie2007StochasticKinetics}.
\end{itemize}

The system state change is then computed in terms of probability by the Chemical Master Equation (CME), which computes the evolution of the probability that the system is in state $\mathbf{X}$ through time. For details about its computation, we refer the reader to the review by El Samad \textit{et al.}~\cite{ElSamad2005StochasticNetworks}. 
An analytical solution of the CME provides the probability density function of the system state $\mathbf{X}(t)$. However, as for deterministic models, the computation of an analytical solution is impossible except for quite simple systems. Therefore, one way to study the behaviour of the system is to construct numerical realizations of the CME. One of the most used method is Gillespie's Stochastic Simulation Algorithm (SSA) \cite{Gillespie1977ExactReactions}. SSA increments the system state at discrete time points, by randomly selecting the next reaction to fire, according to the propensity function of every possible reaction, as well as simulating the event (reaction occurring) time~\cite{Gillespie1977ExactReactions, ElSamad2005StochasticNetworks}. 
Several exact (i.e. simulating every single reaction) alternatives to the original algorithm (the so-called direct method) have been proposed, such as the next-reaction method~\cite{Gibson2000EfficientChannels}, the sorting direct method~\cite{McCollum2006TheBehavior}, and others (see \cite{Gillespie2007StochasticKinetics,Pahle2009BiochemicalApproaches} for a review). However exact algorithms are limited by their computational cost which renders the simulation of large systems intractable. Several approximation methods have been proposed, and have been thoroughly discussed~\cite{Turner2004StochasticReactions, ElSamad2005StochasticNetworks, Karlebach2008ModellingNetworks, Wilkinson2009StochasticSystems}. Approximation simulations such as the very popular tau-leaping method considerably reduce the simulation time, but at the expense of a hardly estimable loss of accuracy~\cite{Wilkinson2009StochasticSystems}. 

Stochastic discrete simulations offer a different perspective on the modelling of regulation compared to the deterministic continuous approach, as they explicitly model the binding of regulator molecules on the target promoter. Taking the example of TFs regulating the expression of a given gene, a stochastic model can represent the binding of regulatory molecules on the promoter of the target gene as follows:

\begin{align*}
\text{Promoter} + \text{TF} &\rightarrow \text{Active\_promoter} \\
\text{Active\_promoter} &\rightarrow \text{Promoter} + \text{TF} \\
\text{Active\_promoter}  &\rightarrow \text{Active\_promoter} + \text{mRNA}
\end{align*}

In this model, a TF must be bound to the promoter for a transcription event to occur. Using this representation, it is easy to represent the different combinations of TFs bound to the promoter, and the transcription rate associated with each state. The possible combinatorial regulation effects can also be explicitly stated. For example, reactions can be added to encode the formation of a regulatory complex from the different TFs and to encode the binding of the complex on the target promoter. 

The advantage of a stochastic model as opposed to its deterministic counterpart is its ability to fit more precisely to the natural variation inherent to biological systems. This biological fluctuation can be crucial for understanding certain systems, as illustrated by El Samad \textit{et al.}~\cite{ElSamad2005StochasticNetworks} that present several examples where a deterministic model fails to correctly predict the system behaviour. As for its deterministic counterpart, the stochastic framework implies a number of hypotheses on the represented system. In particular, it relies on the essential assumption that the system is well-stirred, that is, the molecules are homogeneously spread in the volume. Moreover the simulation of temporal trajectories is computationally heavier as each reaction (in the case of exact simulation algorithms) or group of reactions (for approximate methods) is simulated. This is especially true for a system with a high number of molecules or reactions with large value propensity functions, as both factors imply high firing rates.


\subsubsection{Bridging the gap}

  Starting from a stochastic model and in particular the CME representation of a given system, it is possible by means of simplifying assumptions to obtain the corresponding continous deterministic model. As stated in the tau-leaping approximation of the SSA, under the assumption that the time step $\tau$ in the simulation is small enough so that the propensity functions of the different reactions stay approximately constant during the interval $[t, t+\tau]$, the number of reactions occurring during that time step can be modelled as a Poisson process~\cite{ElSamad2005StochasticNetworks,Higham2008ModelingReactions}. Consequently, it is possible to sample the number of occurrences of each reaction with propensity function $a_j$ from a Poisson law with parameter $a_j \cdot \tau$. Moreover, if the time step $\tau$ is at the same time large enough so that each reaction fires more than once during the interval $[t, t+\tau]$ (usually feasible in systems where the concentration of species is large enough~\cite{ElSamad2005StochasticNetworks}), the system can be represented by a set of stochastic differential equations, termed the Chemical Langevin Equation (CLE)~\cite{Gillespie2000ChemicalEquation, ElSamad2005StochasticNetworks}. In the CLE, the population of each species evolves in a continuous but stochastic manner, with the stochastic variation due to each reaction being proportional to the propensity of the reaction. As a consequence, the number of occurrences of a reaction with a high rate will have a higher variance than that of a reaction with a small reaction rate.

By further assuming that the abundance of each species is high enough, the stochasticity can be neglected, and the system is reduced to the Reaction Rate Equation (RRE)~\cite{ElSamad2005StochasticNetworks, Higham2008ModelingReactions}, that is a set of differential equations:

\begin{align}
\frac{d\mathbf{X}(t)}{dt} = \sum \limits_{j=1}^M v_j \cdot a_j(\mathbf{X}(t))
\end{align}

In this equation, the change per time unit in the concentration of a given species amounts to the sum over all reactions of the change in the species abundance triggered by one firing of the reaction (i.e. $v_j$), weighted by the rate of the reaction (i.e. the probability that the reaction will fire in one time unit, $a_j$). As highlighted by Higham~\cite{Higham2008ModelingReactions}, it is important to keep in mind that the solution of a deterministic model or RRE obtained by simplifying a stochastic model is not equivalent to an average of many numerical realisations of the corresponding stochastic model. It is rather a limit towards which these realisations tend when the different simplifying assumptions are fulfilled. 

This connection between the stochastic and deterministic frameworks has been leveraged in the case of multi-scale systems, which are systems in which both slow and fast reactions and/or both low- and high-abundance species are present \cite{ElSamad2005StochasticNetworks}. 
This situation can be encountered for example when simulating gene expression and metabolic reactions in a single model. On one hand, gene expression is a slow process involving genes that are present in only one or two copies per cell, and TFs whose abundance can be as low as a dozen of molecules only. On the other hand, metabolic reactions are fast enzyme-catalysed processes and involve highly abundant metabolic species. The issue with such systems is that some but not all reactions could be represented by a deterministic process, while the rest requires a stochastic modelling. In such cases, the SSA performs poorly because of fast reactions and highly abundant species which monopolise most of the computational time. On the opposite, deterministic models provide a poor approximation of slow reactions and low abundant species. Several hybrid methods have hence been proposed. Their strategy is to split reactions and/or species in two sets of slow and fast reactions/species, and to use the most appropriate representation to model each set (see~\cite{Pahle2009BiochemicalApproaches} for an overview of existing methods). This principle notably underlies the slow-scale SSA algorithm~\cite{Cao2005TheAlgorithm.}. The interested reader is referred to~\cite{ElSamad2005StochasticNetworks,Pahle2009BiochemicalApproaches} for more details.

\subsection{Model simplifications}

In addition to the mathematical framework they employ, existing simulators of expression data differ by a number of assumptions they make. Indeed, while it is crucial to accurately represent biological processes, our incomplete knowledge about the detailed mechanisms dictates the use of assumptions and simplifications in the models. These simplifications also arise from the desired level of complexity and the need for computational efficiency. \\

\begin{table}[htbp]
{
\centering
\scriptsize
\makegapedcells 
\begin{tabularx}{\textwidth}{|c|c|Y|Y|c|}
		\hline
        \makecell{\textbf{Simulation}\\ \textbf{method}} & \makecell{\textbf{Network}\\ \textbf{topology}} & \makecell{\textbf{Mathematical}\\ \textbf{formalism}} & \makecell{\textbf{Simulated}\\ \textbf{molecules}} & \makecell{\textbf{Simulated}\\ \textbf{reactions}}\\ 
        \hline
        \makecell{Mendes \textit{et al.},\\ 2003 \cite{Mendes2003ArtificialAlgorithms}} & \makecell[l]{\tabitem Random\\ \tabitem Small-world\\ \tabitem Scale-free\\ \tabitem Regular grid} & ODEs & mRNAs & \makecell[l]{\tabitem Transcription (regulated, Hill \\ function) \\ \tabitem mRNA decay (1$^{\text{st}}$ order process)}\\
        \hline
        \makecell{Van den Bulcke\\ \textit{et al.}, 2006 \\ \cite{denBulcke2006SynTReN:Algorithms} \\ - SynTReN} & \makecell{Sampling from\\ source network} & \makecell{ODEs\\(steady-state)} & mRNAs & \makecell[l]{\tabitem Transcription (regulated, Hill \\ and Michaelis-Menten functions) \\ \tabitem mRNA decay (1$^{\text{st}}$ order process)}\\
        \hline
        \makecell{Ribeiro \textit{et al.},\\ 2007  \cite{Ribeiro2007SGNSimulator}\\ - SGNSim} & User-defined & \makecell{Time-delay\\stochastic\\ model} & \makecell{Gene promoters,\\ mRNAs,\\ proteins,\\ RNA \\ polymerase, \\ ribosomes} & \makecell[l]{\tabitem Transcription (different transcri- \\ption rate for each promoter state) \\ \tabitem Translation (1$^{\text{st}}$ order process)\\ \tabitem mRNA and protein decay (1$^{\text{st}}$\\ order processes)}\\
        \hline
        \makecell{Roy \textit{et al.}, 2008\\ \cite{Roy2008ATranscription} \\ - RENCO} & \makecell[l]{\tabitem Scale-free \\ (protein-protein\\ interaction)\\ \tabitem Exponential\\ degree distribution\\ (transcription\\ network)} & ODEs & \makecell{mRNAs,\\proteins} & \makecell[l]{\tabitem Transcription (regulated, Hill\\ function) \\ \tabitem Translation (1$^{\text{st}}$ order process)\\ \tabitem mRNA and protein decay (1$^{\text{st}}$\\ order processes)}\\
        \hline
        \makecell{Di Camillo \textit{et al.},\\ 2009 \cite{DiCamillo2009AAlgorithms} \\- NETSim} & \makecell{Hierarchical \\modular topo-\\logy model} & ODEs & mRNAS & \makecell[l]{\tabitem Transcription (regulated, fuzzy\\ logic) \\ \tabitem mRNA decay (1$^{\text{st}}$ order process)}\\
        \hline
        \makecell{Haynes \textit{et al.},\\ 2009 \cite{Haynes2009BenchmarkingGRENDEL}\\ - GRENDEL} & \makecell{Distinct in- \\and out-degree\\ distribution} & ODEs & \makecell{mRNAs, \\ proteins,\\ environ-\\ment} & \makecell[l]{\tabitem Transcription (regulated, Hill\\ function) \\ \tabitem Translation (1$^{\text{st}}$ order process)\\ \tabitem mRNA and protein decay (1$^{\text{st}}$\\ order processes)}\\
        \hline
        \makecell{Hache \textit{et al.},\\ 2009 \cite{Hache2009GeNGe:Networks}\\ - GeNGe} & \makecell[l]{\tabitem Random\\ \tabitem Scale-free\\ \tabitem Regulatory motifs\\ \tabitem User-defined}  & ODEs & \makecell{mRNAs,\\ proteins,\\ RNA polymerase, \\ ribosomes} & \makecell[l]{\tabitem Transcription (regulated, Hill \\ function) \\ \tabitem Translation (1$^{\text{st}}$ order process)\\ \tabitem mRNA and protein decay (1$^{\text{st}}$\\order processes or Michaelis-\\Menten decays)}\\
        \hline
        \makecell{Schaffter \textit{et al.}, \\2011 \cite{Schaffter2011GeneNetWeaver:Methods}\\ - GeneNetWeaver} & \makecell{Module extraction \\ from source network} & \makecell{Chemical \\ Langevin\\ Equation} & \makecell{mRNAs,\\ proteins} & \makecell[l]{\tabitem Transcription (regulated, Hill\\ function) \\ \tabitem Translation (1$^{\text{st}}$ order process)\\ \tabitem mRNA and protein decay (1$^{\text{st}}$\\ order processes)}\\
        \hline
        \makecell{Pinna \textit{et al.}, \\2011 \cite{Pinna2011SimulatingSysGenSIM}\\ - SysGenSIM} & \makecell[l]{\tabitem Random\\ \tabitem Scale-free\\ \tabitem Random modular\\ \tabitem Modular with\\ exponential in-degree\\ and power law out-\\degree} & ODEs & \makecell{mRNAs,\\ cis- and\\ trans-eQTLs} & \makecell[l]{\tabitem Transcription (regulated, Hill\\ function) \\ \tabitem mRNA decay (1$^{\text{st}}$ order process)}\\
        \hline
        \makecell{Tripathi \textit{et al.}, \\2017 \cite{Tripathi2017SgnesR:Parameters}\\ - sgnesR} & User-defined & \makecell{Time-delay\\stochastic\\ model} & \makecell{Gene promoters,\\ mRNAs,\\ proteins} & \makecell[l]{\tabitem Transcription (different transcri- \\ption rate for each promoter state) \\ \tabitem Translation (1$^{\text{st}}$ order process)\\ \tabitem mRNA and protein decay (1$^{\text{st}}$\\ order processes)}\\
        \hline
\end{tabularx}  
\caption{Overview of existing methods of expression data simulation and their characteristics.}
\label{table::overview}
}  
\end{table}

An important aspect to consider when designing a model is the type of molecules one wishes to represent. Early models were restricted to the simulation of the transcript levels only (see Table~\ref{table::overview}), and used the concentration of mRNAs as a proxy for the activity of their protein product. Such an assumption was justified by the inability to experimentally measure protein concentration~\cite{DiCamillo2009AAlgorithms}. However, as shown earlier in this chapter, a number of regulations occur post-transcriptionally. This certainly impacts protein abundance and/or activity without being reflected at the level of corresponding transcripts, except when the expression of a coding gene is linked to the activity of its corresponding protein via a feedback circuit. In particular, many studies revealed a generally weak not to say poor correlation between transcript and protein profiles \cite{Halbeisen2008Post-transcriptionalPrinciples, Vogel2012InsightsAnalyses}. Such results suggest the need for more realistic models in which proteins are also included as the direct actors of transcription regulation. Some models already include the protein level (see Table \ref{table::overview}). The inclusion of other regulatory molecules, and in particular the noncoding yet very likely regulatory~\cite{Wery2011NoncodingRegulation,Morris2014TheRNA,Holoch2015RNA-mediatedExpression} fraction of the transcriptome could also be an interesting development. In addition, post-transcriptional regulations are traditionally overseen in expression simulation methods. Accounting for them would result in an increased  complexity of the underlying mathematical models, but would pave the way for enhanced realism of simulated data.

Biological processes are not instantaneous. Time delays exists between for example transcription initiation and the release of a fully functional mRNA ready to be translated. Such delays have been mostly ignored, to the exception of SGNSim~\cite{Ribeiro2007SGNSimulator} (implemented in the \texttt{R} package \texttt{sgnesR}~\cite{Tripathi2017SgnesR:Parameters}). These stochastic models use a version of the SSA that is suited for the occurrence of delay in biochemical reactions. This can account for the time required for the transcription of a gene as well as the diffusion of molecules across cellular compartments. Additionally, spatial inhomogeneities can be considered. For example, one might want to include in the model the different cellular compartments, to account for the fact that in eukaryotic cells the synthesis of mRNA occurs in the nucleus, while their translation happens in the cytoplasm. This can be done in a deterministic framework by using partial differential equations~\cite{deJong2002ModelingReview.}.

\subsection{Assigning values for model parameters}

When simulating \textit{in silico} expression data, it is important to carefully choose the values of the different parameters in the model to obtain more plausible data. The initial abundance of each molecular species and the different reaction rates determine the resulting level of expression for each gene. It is crucial to use reasonable values in the range of those estimated from experimental datasets. The same attention must be paid to the kinetic parameters that define the strength and amplitude of regulation. This includes, for example, the Hill coefficients for a deterministic model, or the binding and unbinding rates of the different TFs on the promoter for a stochastic model. This choice is impeded by our limited knowledge about the precise kinetics of gene expression regulation~\cite{denBulcke2006SynTReN:Algorithms}. However a number of experimental results provide insights into the dynamics of the different molecular reactions, at least for some model organisms~\cite{Belle2006QuantificationProteome,Vogel2012InsightsAnalyses}. The global distribution of the lifetime of transcripts and proteins, for example, is starting to be well-characterized across the different domains of life. The order of magnitude of transcription and translation rates are also available. In~\cite{Milo2016CellNumbers}, Milo and Phillips gather relevant quantitative pieces of information about different biological processes, from cell component typical size estimates to the rates of transcription, translation or metabolic reactions. The associated database, BioNumbers~\cite{Milo2009BioNumbersBiology}, allows to search the literature for quantitative properties of biological systems. This is a valuable tool for modellers who seek realistic values for model parameters. Additionally, during the model construction it can be useful to conduct a sensitivity analysis to ensure that slight variations in parameter values do not produce completely different and/or surrealistic system behaviours. 

Despite the increasing availability of quantitative knowledge regarding gene expression, to the best of our knowledge, none of the existing simulation tools presently offer a rigorous justification of the values used in their model. The parameters are usually sampled from large distributions to allow a wide variety of possible dynamical behaviour (e.g. from quasi-linear to step-like regulatory functions) \cite{denBulcke2006SynTReN:Algorithms,DiCamillo2009AAlgorithms}. Alternatively, parameter values can be required as input from the user, as it is the case in~\cite{Ribeiro2007SGNSimulator,Hache2009GeNGe:Networks,Tripathi2017SgnesR:Parameters}. However the choice of values in the model often seem arbitrary~\cite{Mendes2003ArtificialAlgorithms, Roy2008ATranscription, Schaffter2011GeneNetWeaver:Methods, Pinna2011SimulatingSysGenSIM}. An interesting approach has been proposed by Haynes \textit{et al.}~\cite{Haynes2009BenchmarkingGRENDEL}. In their model, the transcription, translation, transcript and protein decay rates are sampled from values experimentally measured for real genes in \textit{S. cerevisiae}. It should be noted however that this limits the validity of the simulations to this organism. Moreover, this approach is only possible for well-characterized model organisms for which abundant and reliable quantitative information is available. 

In conclusion, it is important to anchor the mathematical RNA models in the biological reality not only via the represented molecules and interactions, but also through the quantitative information which is used to simulate the different reactions involved in gene expression.

\subsection{Experimental noise in \textit{in silico} data}

Even though the results of transcriptomics and other omics experiments provide an estimation of transcripts or other molecules level, they do not exactly reflect their precise \textit{in vivo} abundance. Each step of the sample preparation process introduces to some extent bias in the quantitative estimation of the molecular concentrations. Such bias in turn impedes our ability to detect correlation between molecular profiles, or introduces spurious correlations, and needs to be accounted for when developing a reverse engineering approach. Consequently, when assessing the performance of such methods, it is important to test their robustness against increasing level of noise in the data. 

Accordingly, several pipelines of data simulation include a step to add experimental noise in the resulting simulated expression profiles to mimic errors and bias introduced by the used measurement technology~\cite{Mendes2003ArtificialAlgorithms, denBulcke2006SynTReN:Algorithms, Haynes2009BenchmarkingGRENDEL, Hache2009GeNGe:Networks, Schaffter2011GeneNetWeaver:Methods, Pinna2011SimulatingSysGenSIM}. This step is particularly relevant for deterministic models, which produce data deprived of both biological and experimental noise. On the opposite, stochastic models already introduce some kind of variability in the dynamic profiles. The generation of \textit{in silico} experimental noise is often based on models linking the measured intensity obtained with a particular technology (e.g. microarray or RNASeq~\cite{Lowe2017TranscriptomicsTechnologies}) to the true underlying concentration~\cite{Rocke2001AArrays, Irizarry2005Multiple-laboratoryPlatforms, Stolovitzky2007DialogueInference}. 
Alternatively, a simple Gaussian noise can be added to the simulated data to introduce variation in order to blur existing correlations among molecular profiles~\cite{Mendes2003ArtificialAlgorithms,Hache2009GeNGe:Networks,Schaffter2011GeneNetWeaver:Methods,Pinna2011SimulatingSysGenSIM}. Mendes \textit{et al.} proposed to use a Gamma distribution for experimental noise, as microarray data are often found to display a non-Gaussian noise and as the Gamma distribution is not centred around its mean.

\section{Concluding remarks}

Biological systems are characterized by great complexity. From a systems perspective, regulatory networks are shaped according to specific properties that can be described mathematically. From a mechanistic perspective, gene expression is regulated at each step of the lifetime of the different gene products, that are transcripts and proteins. Their synthesis, activity and decay is tightly controlled by a vast array of factors ranging from proteins to noncoding transcripts and small molecules. Additionally, these processes are affected by an inherent stochasticity which induces variability in the molecular profiles. Statistical models provide a rich framework to represent this complexity, and it is now possible to generate \textit{in silico} graphs resembling real networks, or to simulate biologically plausible noisy dynamic expression data. A statistical model must be carefully designed to accurately reflect the underlying processes, as for example determining the list of regulators of a molecule, quantifying the effect of regulatory factors, or assigning a value to the different reaction rates or other parameters. 

In this chapter, we particularly focused on the use of GRN models for the simulation of expression data that can serve as benchmark for the testing of network inference algorithms. Indeed it is important that these simulations provide realistic data to allow researchers to draw meaningful conclusions about the performances of reverse engineering methods. However, beyond the problem of simulating expression data, all the identified regulatory relationships allow the modeller to account for a fine description of the underlying biological mechanisms, when such a level of detail is required.

We know that all models presume to a greater or lesser extent a simplification of the underlying biology. As we have shown in this chapter, most simulation models currently consider transcription regulation only, and exclude noncoding RNAs and possibly even proteins. While these simplifications can be justified by our insufficient knowledge about the processes at play or by computational limitations, it results in inadequate models as they overlook the complexity of gene regulation. We would however like to moderate this desire for increasingly detailed models that could be more indicative of the underlying biological and technical influences in the data. Indeed, they allow a great flexibility in the inferred interactions, but this can become problematic and result in overfitting and in the detection of spurious regulations. The need for a trade-off calls for the use of additional data for the reverse engineering problem as well as advanced statistical tools accounting for the missing information.

\section*{Acknowledgements}

We are very grateful for enriching discussions and suggestions on this manuscript made by Samantha Baldwin and Susan Thomson (Plant and Food Research, Lincoln, New Zealand). We would like to thanks reviewers of this chapter for their useful comments.
MV was partly supported by a visiting professor scholarship from Aix-Marseille University.

\bibliography{Mendeley_Chapter_RNA_regulation}
\bibliographystyle{plain}

\end{document}